\documentclass[10pt,conference,letterpaper]{IEEEtran}

\usepackage{algorithm}
\usepackage[noend]{algpseudocode}
\usepackage{amsmath,amssymb,amsthm}
\usepackage{booktabs}
\usepackage{cite}
\usepackage[hidelinks]{hyperref} 
\usepackage{mathtools}
\usepackage{multirow}
\usepackage[autolanguage]{numprint} \npthousandsep{,}
\usepackage{xcolor}

\newcommand{\A}{\ensuremath{\boldsymbol{A}}}
\renewcommand{\Pr}[1]{\text{Pr}\left[#1\right]}
\newcommand{\Hnorm}{\ensuremath{H_\textrm{norm}}}

\newif\ifcomment
\commenttrue    

\newif\iffigabbrv
\figabbrvtrue    
\newcommand{\figtext}{\iffigabbrv Fig.\else Figure\fi}
\newcommand{\figstext}{\iffigabbrv Figs.\else Figures\fi}

\newif\ifrevise
\newcommand{\revise}[1]{\ifrevise\color{blue} #1 \color{black}\else #1\fi}

\newif\ifshowappendix
\showappendixtrue   
\showappendixfalse  

\IEEEoverridecommandlockouts  

\title{Cutting Through the Noise to\\
Infer Autonomous System Topology}

\author{\IEEEauthorblockN{Kirtus G.\ Leyba\IEEEauthorrefmark{1}, Joshua J.\ Daymude\IEEEauthorrefmark{1}, Jean-Gabriel Young\IEEEauthorrefmark{2}\IEEEauthorrefmark{3},\\
M.\ E.\ J.\ Newman\IEEEauthorrefmark{3}\IEEEauthorrefmark{5}, Jennifer Rexford\IEEEauthorrefmark{4} and Stephanie Forrest\IEEEauthorrefmark{1}\IEEEauthorrefmark{5}}\\
\IEEEauthorblockA{\IEEEauthorrefmark{1}Biodesign Center for Biocomputing, Security and Society\\Arizona State University, Tempe, AZ 85281}
\IEEEauthorblockA{\IEEEauthorrefmark{2}Department of Mathematics and Statistics and Vermont Complex Systems Center\\University of Vermont, Burlington, VT 05405}
\IEEEauthorblockA{\IEEEauthorrefmark{3}Department of Physics and Center for the Study of Complex Systems\\University of Michigan, Ann Arbor, MI 48109}
\IEEEauthorblockA{\IEEEauthorrefmark{4}Department of Computer Science\\Princeton University, Princeton, NJ 08540}
\IEEEauthorblockA{\IEEEauthorrefmark{5}The Santa Fe Institute, Santa Fe, NM, 87501}
\IEEEauthorblockA{Emails: \{kleyba, jdaymude, steph\}@asu.edu, jean-gabriel.young@uvm.edu, mejn@umich.edu, jrex@cs.princeton.edu}\thanks{\copyright~IEEE 2022. Published version to appear at IEEE INFOCOM 2022.}\vspace{-8mm}}

\begin{document}

\maketitle

\begin{abstract}
    The Border Gateway Protocol (BGP) is a distributed protocol that manages interdomain routing without requiring a centralized record of which autonomous systems (ASes) connect to which others.
    Many methods have been devised to infer the AS topology from publicly available BGP data, but none provide a general way to handle the fact that the data are notoriously incomplete and subject to error.
    This paper describes a method for reliably inferring AS-level connectivity in the presence of measurement error using Bayesian statistical inference acting on BGP routing tables from multiple vantage points.
    We employ a novel approach for counting AS adjacency observations in the AS-PATH attribute data from public route collectors, along with a Bayesian algorithm to generate a statistical estimate of the AS-level network.
    Our approach also gives us a way to evaluate the accuracy of existing reconstruction methods and to identify advantageous locations for new route collectors or vantage points.
\end{abstract}

\section{Introduction} \label{sec:intro}

Global Internet routing relies on the Border Gateway Protocol (BGP) to route traffic between Autonomous Systems (ASes).
Obtaining a reliable representation of the AS network is relevant to many applications, including inferring business relationships between ASes~\cite{Dimitropoulos2007-asrelationships,Giotsas2014-complexas,Luckie2013-asrelationships}, identifying politically driven Internet dynamics~\cite{Leyba2019-bordersgateways,Douzet2020-ukraine}, and simulating routing protocols~\cite{Gill2012-modelingquicksand}.
However, there is no central record of the AS network topology; it must be inferred from publicly available routing data, such as those from the RIPE Routing Information Service (RIPE RIS)~\cite{ripencc-ris}
and \revise{RouteViews~\cite{uoregon-routeviews}.}
Existing methods for reconstructing the AS topology from public BGP routing table data typically rely on ad hoc assumptions and heuristics when choosing which edges to include~\cite{CAIDA-asrank,Dimitropoulos2007-asrelationships,Luckie2013-asrelationships} and do not rigorously account for the many sources of error in routing data, including unintentionally misconfigured ASes, intentional traffic manipulation (e.g., path poisoning), and the constant churn arising from changes in AS routing policies.
Inaccuracies arising from these issues can generate incorrect edge relationships and harm related applications~\cite{Jin2019-problink}.

Here we use formal statistical inference methods~\cite{Newman2018-estimatingnetwork,Newman2018-networkstructure} to infer, for each pair of ASes in the network, the probability that an edge exists between them given a set of possibly unreliable observations.
We collect routing tables from public route collectors and obtain edge observations from the AS-PATH attributes included in their routing information bases (RIBs).
By extracting both positive observations of an edge's existence and negative observations of an edge's absence from these paths, we maximize the information available.
We employ a statistical model in which each route collector is treated as an independent mode of observation with its own error rates, and we use expectation-maximization (EM) to estimate these error rates and the probability of existence for each edge in the AS network.
Our main contributions are:
\begin{itemize}
    \item A statistical model of BGP measurement methods as generators of noisy edge observations, which maximizes the available information by efficiently counting both positive and negative edge observations.
    \item A Bayesian inference method for network reconstruction applied to the AS topology, a large-scale, path-vector-based network with noisy route collectors.
    \item \revise{A quantitative, information theoretic comparison of our model to other AS network reconstruction methods.}
    \item An example use case showing how our model can  identify geographic regions of the AS network that would benefit most from additional measurement information (in the form of new route collectors).
\end{itemize}

\section{Background} \label{sec:background}

\subsection{Sources of Noise in BGP Data} \label{subsec:noisesources}

Interdomain routing depends on over \numprint{70000} ASes that apply their own routing policies to select paths to remote destinations.
BGP can exhibit a wide range of complex behaviors caused by the interaction of diverse local routing policies, accidental misconfigurations, intentional attacks, and the protocol's own convergence process.
For our purposes, what matters is that each of these phenomena can lead to errors in observations of AS-level edges in the network topology, which we treat in our model as noise.
We begin by reviewing several sources of such errors.

\paragraph{Local routing policies}

Each AS uses local policies to select, for each destination IP prefix, a preferred path from the options offered by its neighbors.
It also decides whether or not to propagate the chosen path to additional neighbors.
For example, business relationships may cause an AS to favor one path over another, or to decide not to advertise a chosen path to a particular neighbor.
Such policies affect which paths are observed by each route collector and can cause collectors to miss edges that actually exist.

\paragraph{Misconfigurations}

Misconfigured ASes may introduce invalid paths containing nonexistent edges, or they may fail to advertise edges that do exist.
Some misconfigurations are localized to particular vantage points~\cite{Luckie2014-spuriousroutes}.
Others can prompt BGP session resets that cause \revise{many} valid paths to be withdrawn and readvertised in a short period of time~\cite{Cheng2011-tabletransfers}.

\paragraph{Prefix hijacking}

Prefix hijacking attacks occur when an AS announces an IP prefix that belongs to another AS as its own; see~\cite{Cimpanu2019-hijackchina,Saarinen2020-hijacktelstra,Siddiqui2020-hijack} for several recent examples.
Naive hijacking simply routes traffic along a path of valid AS-level edges that leads to the wrong destination for the hijacked prefix.
Since all edges are valid, this would not introduce observational errors at a route collector.
In a stealthier attack, however, an adversary originates a route with a fake ``last hop'' that purportedly connects the adversary's AS to the legitimate destination AS for the prefix.
This can cause a route collector to erroneously observe this final, nonexistent edge.

\paragraph{Path poisoning}

\revise{An AS can poison an advertised path by including additional AS numbers that appear to introduce a loop.}
This misleads upstream ASes into filtering out the route as a result of BGP's built-in loop detection mechanisms.
Path poisoning can be used to manipulate the path that traffic takes to reach certain destinations~\cite{BirgeLee2018-bamboozlebgp} or to protect victims of denial-of-service attacks by blocking unwanted traffic from senders in particular upstream ASes~\cite{Smith2018-defeatingddos}.
For example, an AS $i$ \revise{originating a prefix may poison its path by including ASes $j$ and $k$} to trigger loop detection and subsequent route filtering by ASes $j$ and $k$.
A route collector would then erroneously observe nonexistent edges such as $(i, j)$ and $(j, k)$; it would also miss paths from some vantage points.

\paragraph{Systematic noise}

Beyond these explicit sources of error, the constant churn of path advertisements and withdrawals affects the data.
Each advertisement or withdrawal, if propagated to a route collector, causes fluctuations of the edge observations in the routing tables, and these may not be observed consistently across all vantage points.
Further, an AS may change its preferred routes at any time which will eventually propagate to vantage points, creating additional inconsistencies among route collectors.

\smallskip

Taken together, these sources of error introduce significant noise in route collectors' observations of AS-level edges.
To distinguish signal from noise in BGP data, we turn to the statistical inference methods described in the next subsection.

\subsection{Network Inference} \label{subsec:inference}

The fundamental problem we face is inferring the structure of the AS topology from unreliable path data.
Recent work in network science has developed statistical tools for tackling this problem for a broad range of networks across the sciences~\cite{Newman2018-estimatingnetwork,Newman2018-networkstructure,Le2018-estimatingnetwork,Peixoto2018-reconstructingnetworks,Priebe2015-inferenceerrorgraph,Butts2003-networkinference,Jansen2003-bayesiannetworks,Young2021-reconstructionplant}.
Instead of producing a single ``best guess'' network topology, these methods use generative models to produce, for each pair of nodes in a network, an estimate of the probability that they are connected given the available data.
These estimated networks can then be mined to discover particular patterns and to answer substantive research questions about the underlying system~\cite{Khan2018-uncertaingraphs,Pfeiffer2011-nodecentrality}.

In general, these methods consider a network of $N$ nodes whose structure is represented by an $N \times N$ adjacency matrix $\A$ with elements $A_{ij} = 1$ when there is an edge between nodes $i$ and $j$ and $A_{ij} = 0$ otherwise.
In our context, nodes are ASes and edges are direct links.
A generative model for inferring $\A$ based on the observational data $D$ comprises two parts.
The \textit{data model} captures the probability $\Pr{D | \A, \theta}$ of observing the data $D$ (in our context, the routing tables) given that the ground truth structure of the network is $\A$.
The \textit{network model} assigns a prior probability $\Pr{\A | \rho}$ to the structure~$\A$, specifying how likely any given network structure is before making any observations.
The parameters $\theta$ and $\rho$ capture any additional model parameters (e.g., variances in measurement data), of which there may be many and whose values are initially unknown.

By Bayes' rule, the posterior probability of the unknown quantities $\A$, $\theta$, and $\rho$ given the observed data $D$ is
\begin{equation} \label{eq:fullpost}
    \Pr{\boldsymbol{A}, \theta, \rho | D} = \frac{\Pr{D | \boldsymbol{A}, \theta}\Pr{\boldsymbol{A} | \rho}\Pr{\theta}\Pr{\rho}}{\Pr{D}},
\end{equation}
where $\Pr{\rho}$ and $\Pr{\theta}$ are Bayesian priors on the parameters.
Conversely, the probability of the network structure $\A$ alone given the parameters and observed data is
\begin{equation} \label{eq:networkpost}
    q(\A) = \Pr{\boldsymbol{A} | \theta, \rho, D}
    = \frac{\Pr{\boldsymbol{A}, \theta, \rho | D}}{\Pr{\theta, \rho | D}}.
\end{equation}
After observing data $D$, $\Pr{\boldsymbol{A}, \theta, \rho | D}$ gives the joint probability of the network \textit{and} the parameters while $q(\A)$ gives the probability of the network when we already know the parameters.
Section~\ref{sec:example} motivates the advantages of this generative modeling framework and Section~\ref{sec:methods} details how we apply it to our present problem of inferring the AS topology.

\section{Motivating Example} \label{sec:example}

\begin{figure}
    \centering
    \includegraphics[width=0.9\columnwidth]{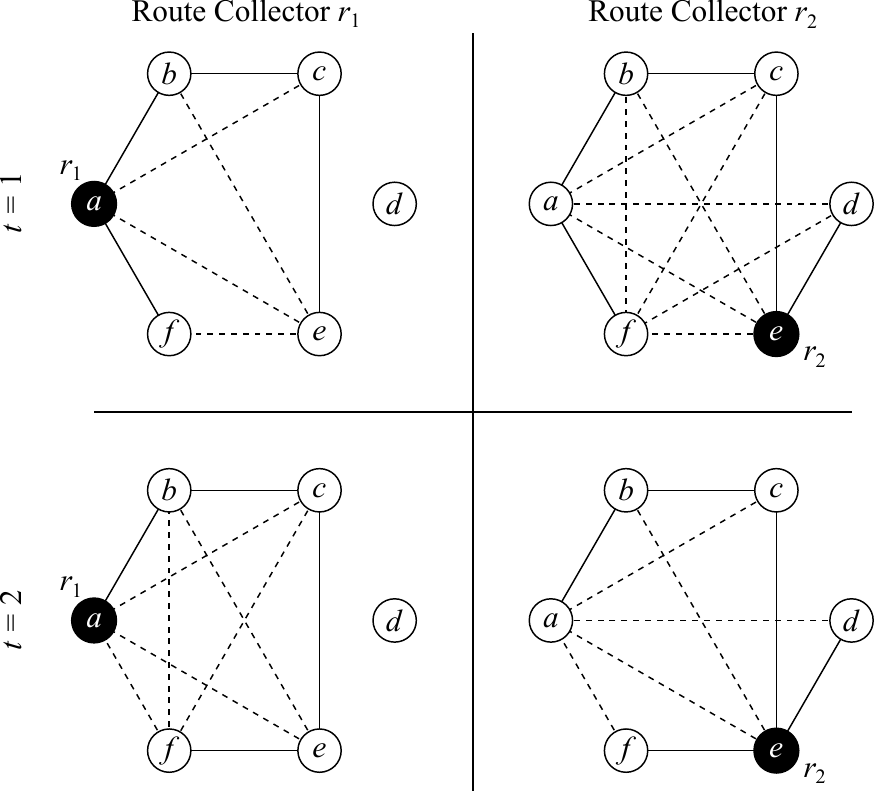}
    \caption{\textbf{Example observations across time and route collectors.}
    Observations of edge existence (solid lines) and absence (dashed lines) gathered by route collectors (black nodes) for an example network of $N = 6$ ASes.
    Edge absences are evidenced by missing connections that would shorten a route among nodes reachable by the route collector (see Section~\ref{subsec:obscount}).
    Due to dynamics and error in BGP data, observations can differ across time periods (rows) and among route collectors (columns).}
    \label{fig:example}
\end{figure}

Suppose we want to infer the topology of an example network of $N = 6$ ASes given the observations in \figtext~\ref{fig:example}.
The observations consist of paths gathered by two route collectors at two time steps.
Usually paths like these are used to extract only ``positive observations'' of the existence of edges---edges that are seen in at least one path---but here we also consider ``negative observations'' of edges' non-existence (see Section~\ref{subsec:obscount} for details).
A naive approach might consider only positive observations by a single route collector in a single time period.
For example, when presented with the top-left network in \figtext~\ref{fig:example}, such an approach would conclude that edges $(a, b)$, $(a, f)$, $(b, c)$, and $(c, e)$ exist while all others do not.
However, extending the observations across multiple time periods shows that the situation is more complex, as edge $(a, f)$ for example is seen in the first time period but not in the second, and vice versa for edge $(e, f)$.
A similar issue arises when comparing the observations of different route collectors: route collector~$r_1$ for example does not observe edge $(d, e)$ in either time period while route collector~$r_2$ observes it in both.

How should these discrepancies be reconciled?
As we describe in the next section, our model assumes each route collector is an independent mode of observation subject to its own rates of observational error.
\revise{For each route collector and time period, we count both the positive (solid) and negative (dashed) edge observations.
We distinguish evidence that an edge does not exist from a mere lack of data---absence of evidence is not evidence of absence.
Utilizing the maximum likelihood parameters of our model, we can then compute, for any pair of ASes in the network, the probability that an edge exists between them.}
For edges such as $(a, b)$, $(b, c)$, and $(c, e)$ in \figtext~\ref{fig:example} that are positively observed by both route collectors in both time periods, this probability will be close to 1.
For edges such as $(a, c)$, $(a, e)$, and $(b, e)$ that are negatively observed by both route collectors in both time periods, the probability will be close to 0.
These probabilities can then be used to evaluate existing BGP network reconstruction methods in terms of our model's certainty (Section~\ref{sec:comparison}).

As with any method for inferring the AS topology from BGP routing data, our method cannot make predictions with high certainty about edges for which there are no observations.
For example, edges $(b, d)$ and $(c, d)$ are never observed as existing or as missing by either route collector in either time period.
For these edges, our method assigns a probability of existence that is proportional to the average frequency with which edges occur in the network as a whole, which is the best guess in the face of complete lack of data.
By locating these low-certainty edges, our method can also suggest network regions that would most benefit from additional measurement in the form of newly constructed route collectors (Section~\ref{sec:newcollectors}).

\section{Methods} \label{sec:methods}

\begin{figure*}[t]
    \centering
    \includegraphics[width=\textwidth]{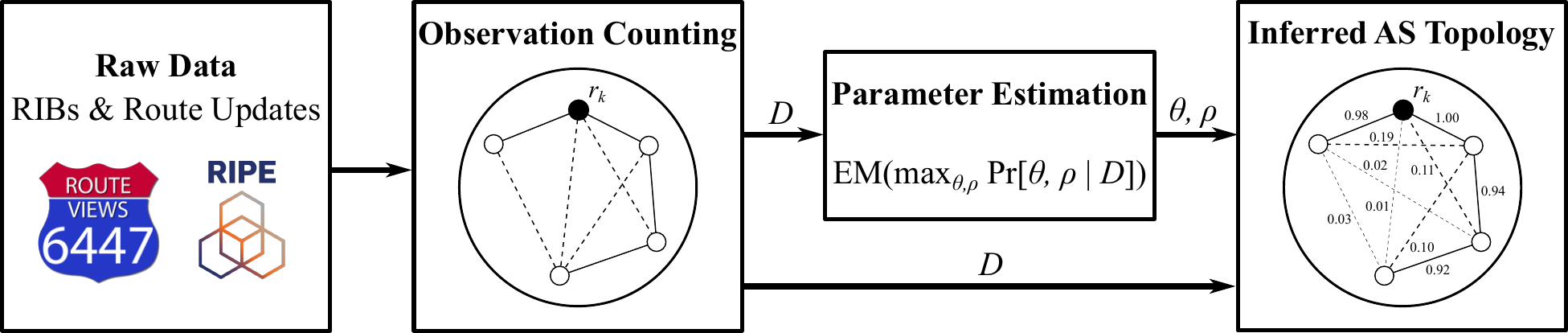}
    \caption{\textbf{Inferring the AS Topology.}
    \textit{Raw data:} BGP routing data are collected as routing information bases (RIBs) and route updates from the RIPE RIS~\cite{ripencc-ris} and RouteViews~\cite{uoregon-routeviews}.
    \textit{Observation counting:} Edge observations are obtained from paths collected by route collectors, e.g., $r_k$.
    Edges in paths are observed positively (solid lines) while absent edges whose existence would create shorter paths from $r_k$ are observed negatively (dashed lines).
    \textit{Parameter estimation:} The route collector observational error rates $\theta$ and the prior $\rho$ are estimated from the data $D$ using expectation-maximization.
    \textit{Inferred AS topology:} The estimated parameters are used to infer, for each pair of ASes, the posterior probability of an edge existing between them.}
    \label{fig:methods}
\end{figure*}

We now detail the components of our statistical inference approach; see \figtext~\ref{fig:methods} for an overview.

\subsection{Data Collection} \label{subsec:data}

The two best known organizations aggregating routing data are the RIPE Routing Information Service (RIPE RIS)~\cite{ripencc-ris} and the University of Oregon's RouteViews Project~\cite{uoregon-routeviews}.
These organizations provide compilations of routes in the form of routing information bases (RIBs) and route updates (both announcements and withdrawals).
We use data from both RIPE RIS and RouteViews, downloaded and parsed using CAIDA's BGPStream~\cite{Orsini2016-bgpstream}.
Our dataset includes observations from the paths of all 45 publicly available route collectors over five 8-hour time periods\footnote{We chose 8-hour time periods since every route collector is guaranteed to dump at least one RIB in this duration.} beginning on May 1st, 2021.

\subsection{Observation Counting} \label{subsec:obscount}

The raw routing data are gathered over a series of $T$ discrete time periods by $M$ route collectors denoted $r_1, \dots, r_M$.
During a time period $t$, route collector $r_k$ gathers a set of paths $P_{k,t}$ from its associated vantage points detailing the sequence of AS nodes included in advertised AS-PATHs for~$r_k$.
By taking the union over the paths in $P_{k,t}$, we obtain a connected graph $G_{k,t}$ representing what route collector $r_k$ observes about the topology in time period $t$, omitting unreachable ASes.

As mentioned before, a key difference between our approach and earlier work is our use of evidence about both the existence and the absence of edges. 
We say an edge between ASes $i$ and $j$ is \textit{positively observed} by route collector $r_k$ in time period $t$ if $(i, j) \in G_{k,t}$.
Note that such observations are binary: we only count whether or not an edge is present in $G_{k,t}$, not how many distinct paths in $P_{k,t}$ it was present in.\footnote{Counting the number of paths an edge appears in instead of using a binary representation would bias the inference. Route collectors with more frequent updates generate more observations such that edges closer to those route collectors would be over-represented in our data.}

Negative observations require a different counting method in order to differentiate between edges that are absent and those that a route collector simply has no information about.
Here we appeal to standard BGP routing policies which generally prefer to select and propagate shorter routes over longer ones~\cite{Anwar2015-routinginthewild,Caesar2005-bgprouting}.
For a route collector $r_k$ and an AS $i$, let $d_t(r_k, i)$ denote the number of edges in the shortest $(r_k, i)$-path in the graph $G_{k,t}$.
An edge between ASes $i$ and $j$ is said to be \textit{negatively observed} by route collector $r_k$ in time period $t$ if $|d_t(r_k, i) - d_t(r_k, j)| \geq 2$ and $(i, j) \not\in G_{k,t}$; i.e.,~if the existence of $(i, j)$ would have provided a shorter path from $r_k$ to either $i$ or $j$ and yet that shorter path was not in the data.
As with positive observations, negative observations are binary.

We recognize the possibility that such an inference of the non-existence of an edge can be wrong.  Observations of an edge's absence, like any observation, can be in error---this is the central premise of our work.
For instance, not all announced paths of the AS network need in fact be as short as possible, meaning that we may erroneously infer that certain edges are absent when they are not.
As we discuss in the following section, the extent of such errors is estimated along with those from other error sources by our inference method.

Our observation counting algorithm (Algorithm~\ref{alg:obscount}) takes as input the graphs $G_{k,t}$ represented as (sparse) adjacency matrices for each route collector $r_k \in \{r_1, \ldots, r_M\}$ and each time period $t \in \{1, \ldots, T\}$.
It produces a mapping of each AS pair $(i, j)$ to its \textit{observation vector} $[E_{ij}^{(1)}, F_{ij}^{(1)}, \ldots, E_{ij}^{(M)}, F_{ij}^{(M)}]$ where $E_{ij}^{(k)}$ (resp., $F_{ij}^{(k)}$) is the number of time periods in which $(i, j)$ was positively (resp., negatively) observed by route collector $r_k$.
Algorithm~\ref{alg:obscount} is presented as a sequential algorithm for clarity; we discuss its parallel implementation that efficiently processes the large-scale network of $N \approx \numprint{73000}$ ASes (or roughly \numprint{2.66e9} AS pairs) in Section~\ref{subsec:optimize}.

\begin{algorithm}[t]
    \caption{Observation Counting} \label{alg:obscount}
    \begin{algorithmic}[1]
        \Statex \textbf{Input:} Graphs $G_{k,t}$ for route collectors $r_k \in \{r_1, \ldots, r_M\}$ and time periods $t \in \{1, \ldots, T\}$.
        \Statex \textbf{Output:} Mapping $D : $ AS pairs $\to$ observation vectors.
        \Statex \textbf{Assumes:} A breadth-first search function \Call{BFS}{$G$, $v$} that returns a list of distances from $v$ to all other nodes in $G$.
        \State Initialize $D_{ij} \gets [0, \ldots, 0]$ for all AS pairs $(i, j) \in V^2$.
        \For {each $r_k \in \{r_1, \ldots, r_M\}$}
            \For {each $t \in \{1, \ldots, T\}$}
                \State Let $d_{k,t} \gets$ \Call{BFS}{$G_{k,t}$, $r_k$}.
            \EndFor
        \EndFor
        \For {each AS pair $(i, j) \in V^2$} \label{alg:obscount:pairloop}
            \For {each $r_k \in \{r_1, \ldots, r_M\}$}
                \For {each $t \in \{1, \ldots, T\}$}
                    \If {$(i, j) \in G_{k,t}$}
                        \State Update $E_{ij}^{(k)} \gets E_{ij}^{(k)} + 1$.
                    \ElsIf {$|d_{k,t}(r_k, i) - d_{k,t}(r_k, j)| \geq 2$}
                        \State Update $F_{ij}^{(k)} \gets F_{ij}^{(k)} + 1$.
                    \EndIf
                \EndFor
            \EndFor
        \EndFor
        \State \Return $D$.
    \end{algorithmic}
\end{algorithm}

\subsection{The Data and Network Models} \label{subsec:model}

As discussed earlier, BGP has many sources of error that can introduce inaccuracy into our observation counts.
In terms of measurement error, the observations can be broken down into four types: true positives and negatives (i.e.,~correct observations of the presence and absence of edges) and false positives and negatives (observations of the presence of nonexistent edges and of the absence of edges that do exist).
We parameterize these error rates by the true positive rate $\alpha$ and the false positive rate $\beta$.
No additional parameters are needed for the true and false negative rates---they are equal to $1 - \beta$ and $1 - \alpha$, respectively.
In practice, not all route collectors may be equally trustworthy, so we assign each route collector $r_k$ its own true and false positive rates $\alpha_k$ and $\beta_k$.

Letting $\boldsymbol{\alpha} = (\alpha_1, \ldots, \alpha_M)$ and $\boldsymbol{\beta} = (\beta_1, \ldots, \beta_M)$, the probability of the $M$ route collectors making observations $D_{ij} = [E_{ij}^{(1)}, F_{ij}^{(1)}, \ldots, E_{ij}^{(M)}, F_{ij}^{(M)}]$ of an edge between ASes $i$ and $j$ can be written as
\begin{equation} \label{eq:dataedgepos}
    \Pr{D_{ij} | A_{ij} = 1, \boldsymbol{\alpha}} = \prod_{k=1}^M \alpha_k^{E_{ij}^{(k)}} (1 - \alpha_k)^{F_{ij}^{(k)}}
\end{equation}
or
\begin{equation} \label{eq:dataedgeneg}
    \Pr{D_{ij} | A_{ij} = 0, \boldsymbol{\beta}} = \prod_{k=1}^M \beta_k^{E_{ij}^{(k)}} (1 - \beta_k)^{F_{ij}^{(k)}},
\end{equation}
depending on whether or not the edge truly exists in $\A$.
Modeling the measurements of different edges as independent conditioned on the true structure of the network, the data model capturing the probability of observing $D$ given the network structure $\A$ and parameters $\theta = (\boldsymbol{\alpha}, \boldsymbol{\beta})$ is given by
\begin{equation} \label{eq:datamodel}
    \begin{aligned}
        \lefteqn{\Pr{D | \A, \theta} =} \\
        & & \prod_{i < j} \Pr{D_{ij} | A_{ij} = 1, \boldsymbol{\alpha}}^{A_{ij}} \Pr{D_{ij} | A_{ij} = 0, \boldsymbol{\beta}}^{1 - A_{ij}}.
    \end{aligned}
\end{equation}

We use a minimal network model $\Pr{\A | \rho}$ that assumes all network edges have the same prior probability $\rho$ of existing (i.e., $\rho$ is analogous to sparsity), yielding
\begin{equation} \label{eq:networkmodel}
    \Pr{\A | \rho} = \prod_{i < j} \rho^{A_{ij}} (1 - \rho)^{1 - A_{ij}}.
\end{equation}
We also assign uniform priors in the interval $[0, 1]$ for the rates $\boldsymbol{\alpha}$ and $\boldsymbol{\beta}$ and for the prior $\rho$, meaning that $\Pr{\theta} = \Pr{\rho} = 1$.

\subsection{Inferring the AS Topology} \label{subsec:algorithm}

We can use the model above to estimate the structure of the AS network using established methods of statistical inference~\cite{Newman2018-estimatingnetwork}.
Substituting the definitions of the data and network models (Eqs.~\ref{eq:datamodel} and~\ref{eq:networkmodel}) into Eq.~\ref{eq:networkpost}, we find that the probability of the network having structure $\A$ given the parameters and observational data can be written in the simple product form:
\begin{equation} \label{eq:networkpostQ}
    q(\A) = \Pr{\A | \theta, \rho, D}
    = \prod_{i < j} Q_{ij}^{A_{ij}} (1 - Q_{ij})^{1 - A_{ij}},
\end{equation}
where
\begin{equation} \label{eq:edgepost}
    \begin{aligned} 
        \lefteqn{Q_{ij}(\theta, \rho) = \Pr{A_{ij} = 1 | \theta, \rho, D} =} \\
        & & \frac{\rho\, \Pr{D_{ij} | A_{ij} = 1, \boldsymbol{\alpha}}}{\rho\, \Pr{D_{ij} | A_{ij} = 1, \boldsymbol{\alpha}} + (1 - \rho) \Pr{D_{ij} | A_{ij} = 0, \boldsymbol{\beta}}}
    \end{aligned}
\end{equation}
is the posterior probability that an edge exists between ASes $i$ and $j$ given the parameters and
\ifshowappendix
\else
data.
\fi

In practice, we do not know a priori the values of the parameters $\theta = (\boldsymbol{\alpha}, \boldsymbol{\beta})$ and $\rho$, so they too must be estimated from the data.
We employ a standard EM algorithm for this task: the most likely parameters $(\theta^*, \rho^*)$ given the data $D$ are those that maximize
\begin{align} \label{eq:parampost}
    \lefteqn{\Pr{\theta, \rho | D} = \sum_{\A} \Pr{\A, \theta, \rho | D} =} \\
    & & \prod_{i < j} \frac{\rho\, \Pr{D_{ij} | A_{ij} = 1, \boldsymbol{\alpha}} + (1 - \rho) \Pr{D_{ij} | A_{ij} = 0, \boldsymbol{\beta}}}{\Pr{D}}, \nonumber
\end{align}
where we have used the fact that $\Pr{\theta} = \Pr{\rho} = 1$.
The value of $Q_{ij}(\theta^*, \rho^*)$ then gives us the posterior probability of an edge $(i, j)$ existing in the AS topology.
We make extensive use of these probabilities in the following sections.

\section{Model Validation} \label{sec:validation}

There is no ground truth dataset for the AS structure against which to validate the predictions of our analysis, but we can use statistical tests to validate our method's consistency and determine if the model is a good fit to the data.
First, we can evaluate whether our model draws sensible conclusions and is capable of accurately representing the observational data using the common Bayesian validation technique called a posterior predictive check~\cite{Young2021-bayesian}.
In essence, this technique works by taking the fitted model, using it to generate synthetic data, and then comparing the results to the real measured data.
Specifically, we use the model to generate a synthetic network in which each edge $(i,j)$ appears with its inferred probability~$Q_{ij}$, then we generate simulated observations of these edges with errors introduced according to the fitted values of the error parameters.
We count the positive observations of each edge and compare them to the counts appearing in the real data.  
These two datasets will be similar if the model captures the structure present in the real data.

\begin{figure}
    \centering
    \includegraphics[width=0.85\columnwidth]{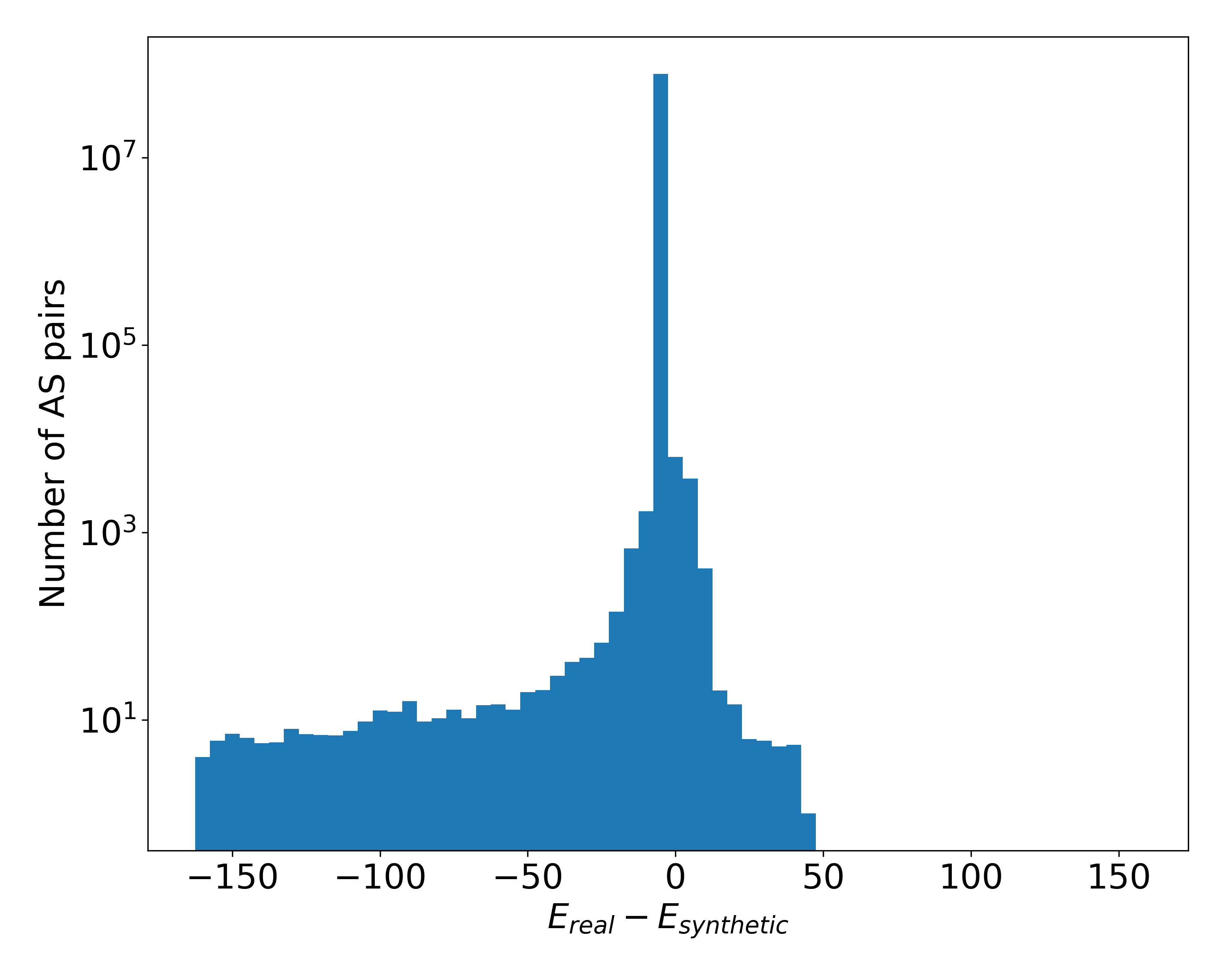}
    \caption{\textbf{Posterior predictive check.}
    Distribution of differences between the number of positive observations per edge in real measurement data and in five synthetic datasets generated from the observational error rates inferred by our model.
    The histogram has 64 bins each of size 5.}
    \label{fig:posteriorcheck}
\end{figure}

\figtext~\ref{fig:posteriorcheck} shows the distribution of the difference between the numbers of positive observations in real and synthetic data over all edges.
We find that the vast majority of AS pairs have exactly the same number of positive observations in the synthetic data as in the real data, and that the number of AS pairs with different numbers of positive observations sharply decreases as that difference increases.
The longer negative tail in \figtext~\ref{fig:posteriorcheck} indicates that the model occasionally predicts too many positive observations, but these differences are minor (note the logarithmic scale for the vertical axis), so we deem the model sufficiently accurate to be put to use.

A complementary check on our method is to ask how confident we are in the method's predictions.
In effect, what are the ``error bars'' on our inferred network structure?
As in any scientific measurement, we want to say not only what we believe the structure of the network to be but also how sure we are of that structure.

\begin{figure}
    \centering
    \includegraphics[width=0.85\columnwidth]{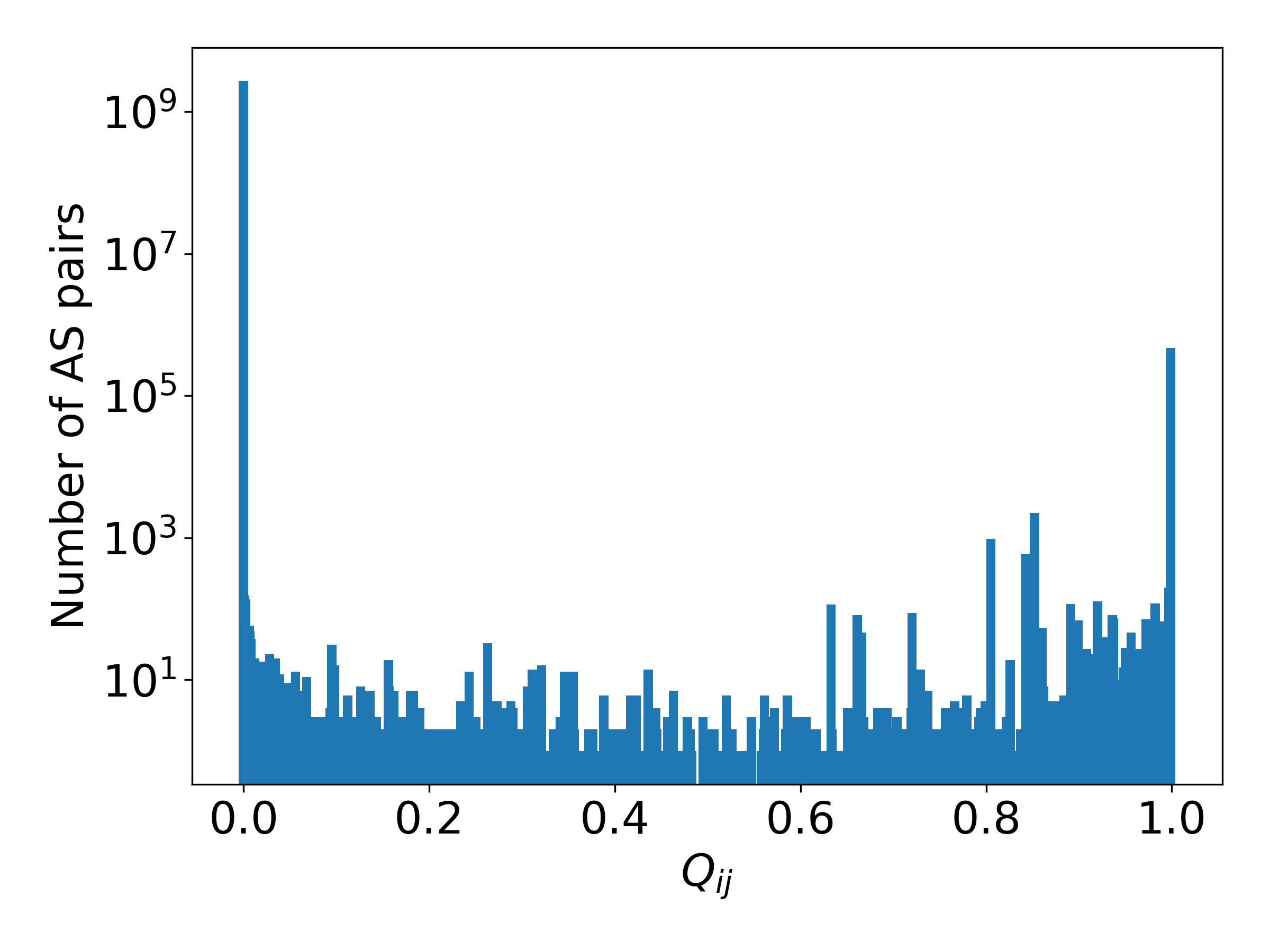}
    \caption{\textbf{Edge posterior probabilities.}
    Distribution of posterior edge probabilities, $Q_{ij}$, inferred by the model from the May 2021 dataset (see Section~\ref{subsec:data}).}
    \label{fig:posteriordist}
\end{figure}

Our confidence is captured by the edge probabilities~$Q_{ij}$.
If $Q_{ij}$ is 1 or 0, then we are certain that the edge between $i$ and $j$ does or does not exist, respectively.
If $Q_{ij} = 0.5$ then the edge is equally likely to exist or not and we are maximally uncertain.
\figtext~\ref{fig:posteriordist} shows a histogram of $Q_{ij}$ values for our May 2021 dataset.
The vast majority of AS pairs
have very small probability of being connected, reflecting the sparsity of the network (note, again, the logarithmic vertical axis).
Of the approximately $\numprint{2.705e9}$ AS pairs, over $99.9\%$ are predicted to exist with probability $Q_{ij} < 0.1$ while fewer than $0.0002\%$ have intermediate probabilities $Q_{ij} \in [0.1, 0.9]$.
Among edges that are predicted to exist, most have a $Q_{ij}$ close to~1.
Thus, the model has high confidence about most predictions---edges either almost certainly exist or almost certainly do not exist.

We can quantify our confidence further by computing the entropy of the edges.
For an edge with probability~$Q_{ij}$, the entropy is defined as
\begin{equation} \label{eq:edgeentropy}
    H(Q_{ij}) = -Q_{ij}\log Q_{ij} - (1 - Q_{ij})\log(1 - Q_{ij}).
\end{equation}
The entropy measures the \emph{un}certainty (or lack of confidence) in our conclusion.
It is zero when $Q_{ij}$ is 1 or 0---meaning we have complete certainty that the edge does or does not exist---and takes a maximal value of $\log2$ when $Q_{ij} = 0.5$ and the edge is equally likely to exist or not.

Since our approach models edges as conditionally independent, the total entropy of the entire inferred network is simply the sum over the set~$V^2$ of all node pairs $\sum_{(i,j)\in V^2} H(Q_{ij})$.
For practical purposes it is useful to normalize this measure to fall on a meaningful scale.
If we have no data about the network at all, then our method will conclude that all edges exist with the prior probability~$\rho$, and hence that the entropy of the network is $H(\rho) = -\rho \log\rho - (1-\rho)\log(1-\rho)$.
We can thus define a normalized entropy per node pair:
\begin{align} \label{eq:normentropy}
    \Hnorm &= \frac{1}{|V^2|} \sum_{(i,j) \in V^2} \frac{H(Q_{ij})}{H(\rho)} \\
    &= \frac{1}{|V^2|} \sum_{(i,j)\in V^2} \frac{Q_{ij}\log Q_{ij} + (1 - Q_{ij})\log(1 - Q_{ij})}{\rho\log\rho + (1 - \rho)\log(1 - \rho)}, \nonumber
\end{align}
which is equal to 0 when we are entirely confident of the inferred network and 1 when we know nothing about it.

\begin{figure}
    \centering
    \includegraphics[width=0.85\columnwidth]{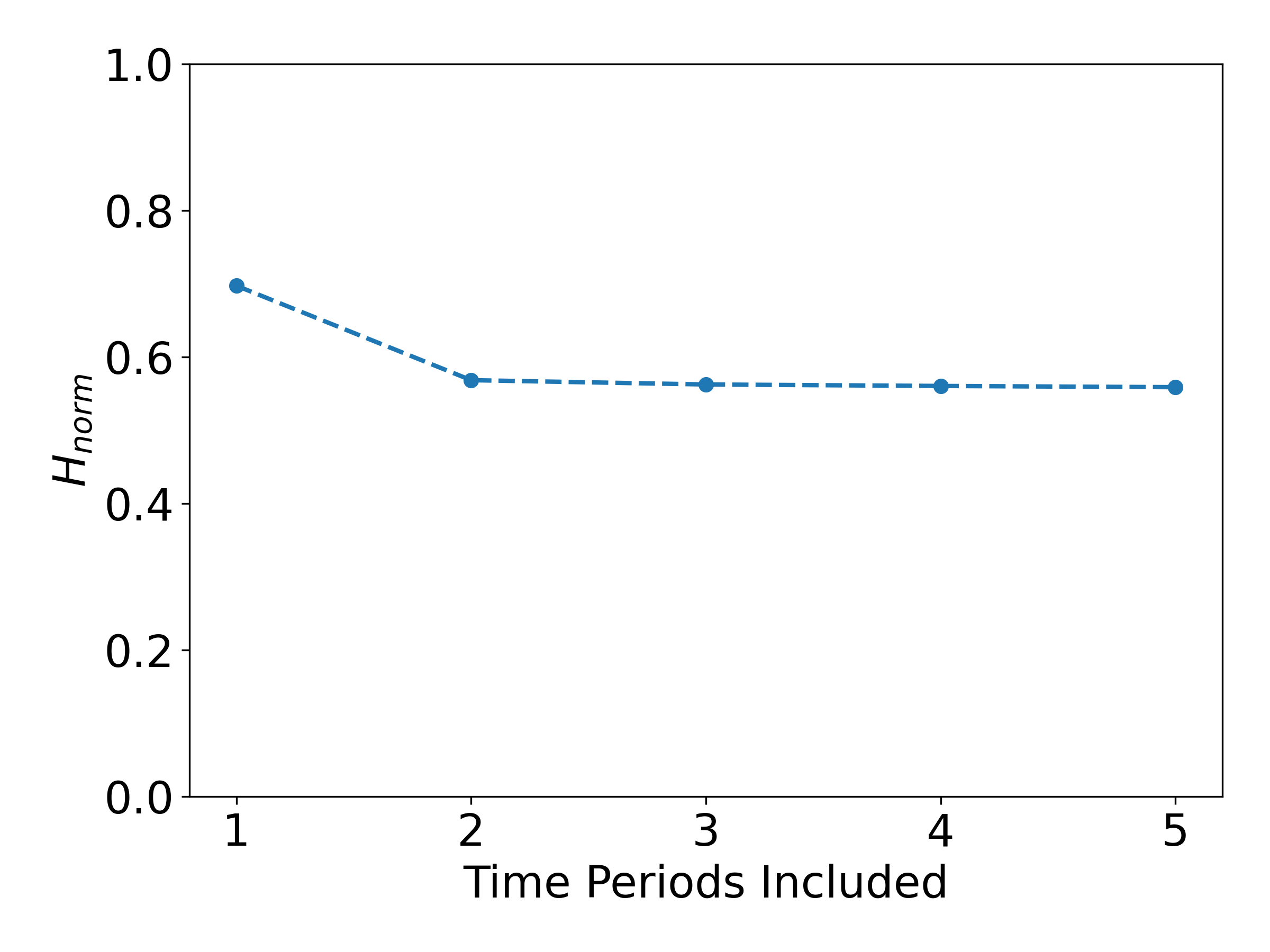}
    \caption{\textbf{Normalized entropy by number of included time periods.}
    Each time period represents 8 hours, the longest time period between RIB dumps for all route collectors available.
    Normalized entropy is measured by $\Hnorm$, defined in Eq.~\ref{eq:normentropy}.}
    \label{fig:timeperiods}
\end{figure}

\figtext~\ref{fig:timeperiods} shows the value of this normalized entropy for our model's representation of the AS network based on data from a single time period and from multiple time periods.
As the figure shows, the model's confidence increases when using multiple time periods instead of just one, but then quickly saturates around $\Hnorm \approx$ 0.56.
Intuitively, this means that once the inferred edge probabilities are known, we could exactly describe the network structure with only 56\% of the information we would have needed if we had no knowledge.
This offers some justification for our use of $T = 5$ time periods of data---a larger number of time periods will give only a marginal improvement in the results, at best.

\section{Comparing Network Reconstructions} \label{sec:comparison}

Next, we consider several AS network reconstructions, comparing the edges they identify to the edge posterior probabilities inferred by our model.
Intuitively, we desire a reconstruction that includes edges that likely exist and excludes those that likely do not exist.
Most earlier methods, however, use heuristics such as loop filtering and clique detection to decide which edges to include.
We consider the following reconstruction methods:
\begin{itemize}
    \item \textit{Naive}. Includes any edge appearing in an advertised path collected by at least one route collector during the collection period.
    
    \item \textit{CAIDA AS Relationships}~\cite{Dimitropoulos2007-asrelationships, Luckie2013-asrelationships}.
    Includes edges corresponding to inferred business relationships, which are obtained by collecting one RIB per route collector per day for the first five days of a month, compressing path padding, removing AS sets, filtering out paths containing loops and those that would separate cliques, and finally, filtering out unassigned ASes and IXP route servers.
    We evaluate two versions of this method: one that includes multi-lateral peering (MLP) data and one that does not.
    
    \item \textit{Threshold $\tau$}.
    Includes any edge $(i,j)$ with $Q_{ij} > \tau$, where $Q_{ij}$ is the edge probability inferred by our model.
    We generate three such topologies from the inferred edge probabilities: one that includes all but the least likely edges ($\tau = 0.1$), another that includes the majority of likely edges ($\tau = 0.5$), and finally one that includes only the most likely edges ($\tau = 0.9$).
\end{itemize}


\begin{table}
    \centering
    \caption{Evaluation of various network reconstructions}
    \label{tab:reconstructions}
    \begin{tabular}{cllll}
        \toprule
        & Method & $\log q(\A)$ & $\text{prec}(\A)$ & $\text{rec}(\A)$ \\
        \midrule
        \parbox[t]{2mm}{\multirow{3}{*}{\rotatebox[origin=c]{90}{IPv4}}} & Naive & \numprint{-9.213e5} & 0.961 & 0.839 \\
        & CAIDA AS Rel. & \numprint{-3.043e6} & 0.855 & 0.638 \\
        & CAIDA AS Rel.\ (MLP) & \numprint{-4.104e6} & 0.690 & 0.641 \\
        \midrule
        \parbox[t]{2mm}{\multirow{4}{*}{\rotatebox[origin=c]{90}{IPv4/6}}} & Naive & \numprint{-3.303e5} & 0.959 & 0.916 \\
        & Threshold $\tau = 0.1$ & \numprint{-4.547e4} & 0.997 & 0.916 \\
        & Threshold $\tau = 0.5$ & \numprint{-4.492e4} & 0.998 & 0.916 \\
        & Threshold $\tau = 0.9$ & \numprint{-5.236e4} & 0.999 & 0.908 \\
        \bottomrule
    \end{tabular}
\end{table}

We quantify reconstruction quality using the standard log-probability of the adjacency matrix~$\A$ under the network posterior $q(\A)$ of Eq.~\ref{eq:networkpostQ}:
\begin{equation} \label{eq:lognetworkpost}
    \log q(\A) = \sum_{i < j} \bigr[A_{ij}\log Q_{ij} + (1 - A_{ij})\log(1 - Q_{ij}) \bigl],
\end{equation}
which is nonpositive and becomes increasingly negative as the probability of $\A$ being the true structure decreases.
We also report the precision (fraction of edges included in $\A$ that exist according to our model) and recall (fraction of edges that exist according to our model that are included in $\A$).
These quantities are given by
\begin{equation}
    \text{prec}(\A) = \frac{\sum_{i < j} A_{ij} Q_{ij}}{\sum_{i < j} A_{ij}}, \quad
    \text{rec}(\A) = \frac{\sum_{i < j} A_{ij} Q_{ij}}{\sum_{i < j} Q_{ij}}.
\end{equation}

Table~\ref{tab:reconstructions} shows our results.
We find that including both IPv4 and IPv6 routes in a naive reconstruction slightly degrades precision but significantly improves network probability and recall over the naive reconstruction that uses only IPv4 routes.
This highlights the importance of including IPv6 data in network reconstructions; in particular, the difference in recall indicates that roughly $7.7\%$ of the edges our model believes exist are only present in IPv6 routes.

The CAIDA AS Relationships reconstructions, both with and without MLP data, have lower probability, precision, and recall than those produced by the naive and threshold methods.
\revise{From the perspective of our model---and, by extension, the public routing table data its inference is based on---this suggests that the heuristics used in the CAIDA reconstructions include some unlikely edges and exclude some likely edges.
Although the CAIDA reconstructions might erroneously include non-existent edges, lower precision could also be explained by inclusion of real edges that are missing from our model's data.
For example, the low precision of the CAIDA MLP reconstruction is likely explained in part by its inclusion of peering relationships between ASes and Internet exchange points (IXPs), which are not detected by conventional parsing of public routing tables.
Thus, our results do not indicate that CAIDA reconstructions are strictly wrong,
but rather show the importance of data, i.e., the decisions about which edges to include or exclude, independently of the inference method.}

\section{Identifying Sites for New Route Collectors} \label{sec:newcollectors}

Although our method extracts maximum information from available observational data, by definition it cannot have high confidence about AS pairs that have conflicting observations or no observations at all.
We leverage this fact to identify regions of the network where additional measurement in the form of new route collectors and vantage points would be most beneficial, addressing a key problem for both BGP network operators and Internet measurement researchers.

\begin{figure}
    \centering
    \includegraphics[width=0.85\columnwidth]{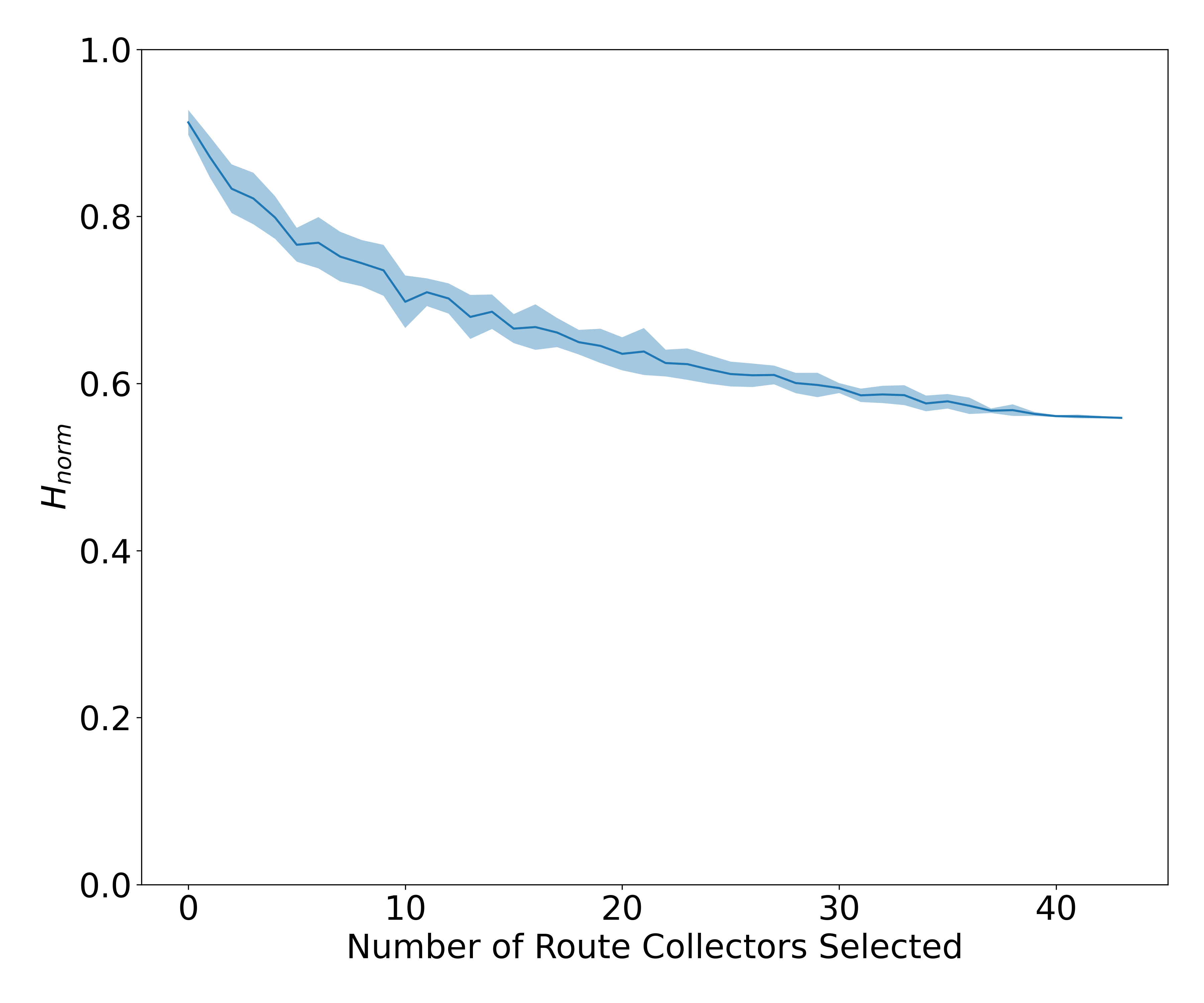}
    \caption{\textbf{Normalized entropy by number of included route connectors.}
    The mean (solid line) and standard deviation (shaded area) of normalized entropy (Eq.~\ref{eq:normentropy}) over 10 randomly shuffled orderings of the 45 route collectors.}
    \label{fig:routecollectors}
\end{figure}

We first establish that including additional route collectors increases our method's certainty in its inference.
Using the May 2021 dataset, we randomize the order of the 45 route collectors ten times and then, for each randomization, fit our model to the data from the first $k$ route collectors, for all $k \in \{1, \ldots, 45\}$.
\figtext~\ref{fig:routecollectors} shows that the normalized entropy~$\Hnorm$ decreases as more route collectors are added, reflecting the model's increasing certainty.
However, the normalized entropy of $\Hnorm \approx$ 0.56 obtained with all 45 route collectors indicates that there is still significant uncertainty in the data---uncertainty that would be further reduced by deploying new route collectors in areas of poor measurement.

To identify under-measured network regions, we use an entropy measure that captures the model's uncertainty about the connectivity of a given AS.
Specifically, for AS~$i$ we define
\begin{equation} \label{eq:nodeentropy}
    H(i) = \sum_{j(\ne i)} H(Q_{ij}),
\end{equation}
where $H(Q_{ij})$ is the edge entropy defined in Eq.~\ref{eq:edgeentropy}. 
When the model is certain about the presence or absence of edges incident to AS~$i$, $H(i)$ is small; otherwise, $H(i)$ is large.  

\begin{figure*}
    \centering
    \includegraphics[width=\textwidth]{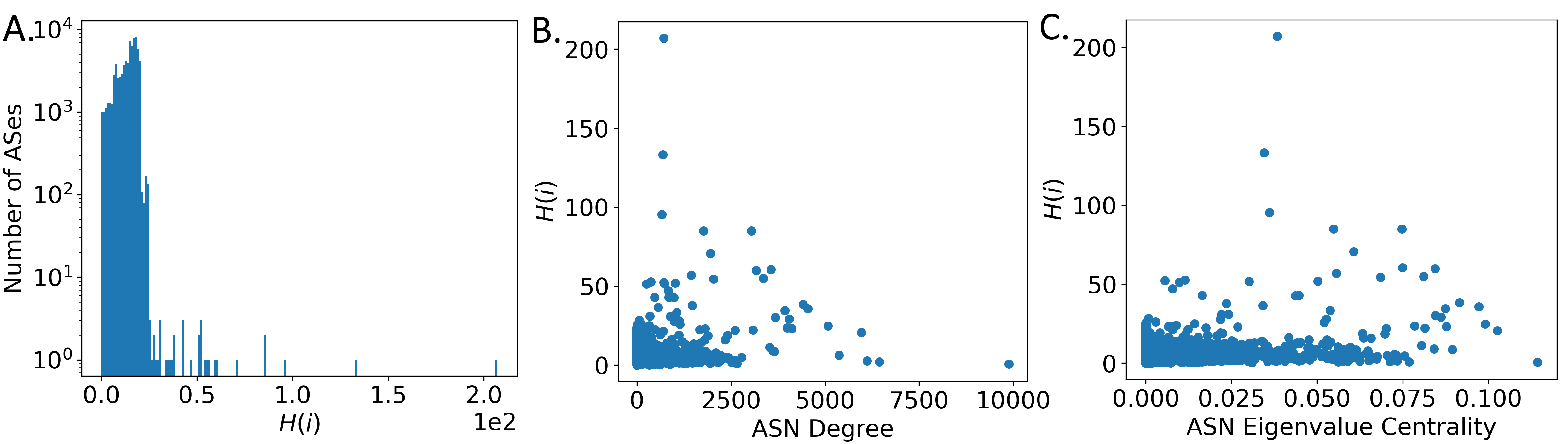}
    \caption{\textbf{AS entropy by various network statistics.}
    \textbf{(A)} Distribution of AS entropy $H(i)$, defined in Eq.~\ref{eq:nodeentropy}.
    \textbf{(B)} Scatter plot of AS entropy $H(i)$ against the degree of an AS in the graph formed from the union of all positive edge observations.
    \textbf{(C)} Scatter plot of AS entropy $H(i)$ against the eigenvalue centrality of an AS in the graph formed from the union of all positive edge observations.}
    \label{fig:nodeentropy}
\end{figure*}

\figtext~\ref{fig:nodeentropy} shows the distribution of this measure as a function of several network statistics.
\figtext~\ref{fig:nodeentropy}A shows that our inference is well-informed about the vast majority of ASes, but the long tail points to outliers about which the model lacks connectivity information.
\figstext~\ref{fig:nodeentropy}B and~\ref{fig:nodeentropy}C show that $H(i)$ generally decreases as AS connectivity increases (as quantified by degree and eigenvector centrality), as expected.
Again, the outliers are interesting; although these ASes are well-connected in the AS network, the model suggests that they lack definitive connectivity information.

\begin{figure}
    \centering
    \includegraphics[width=\columnwidth]{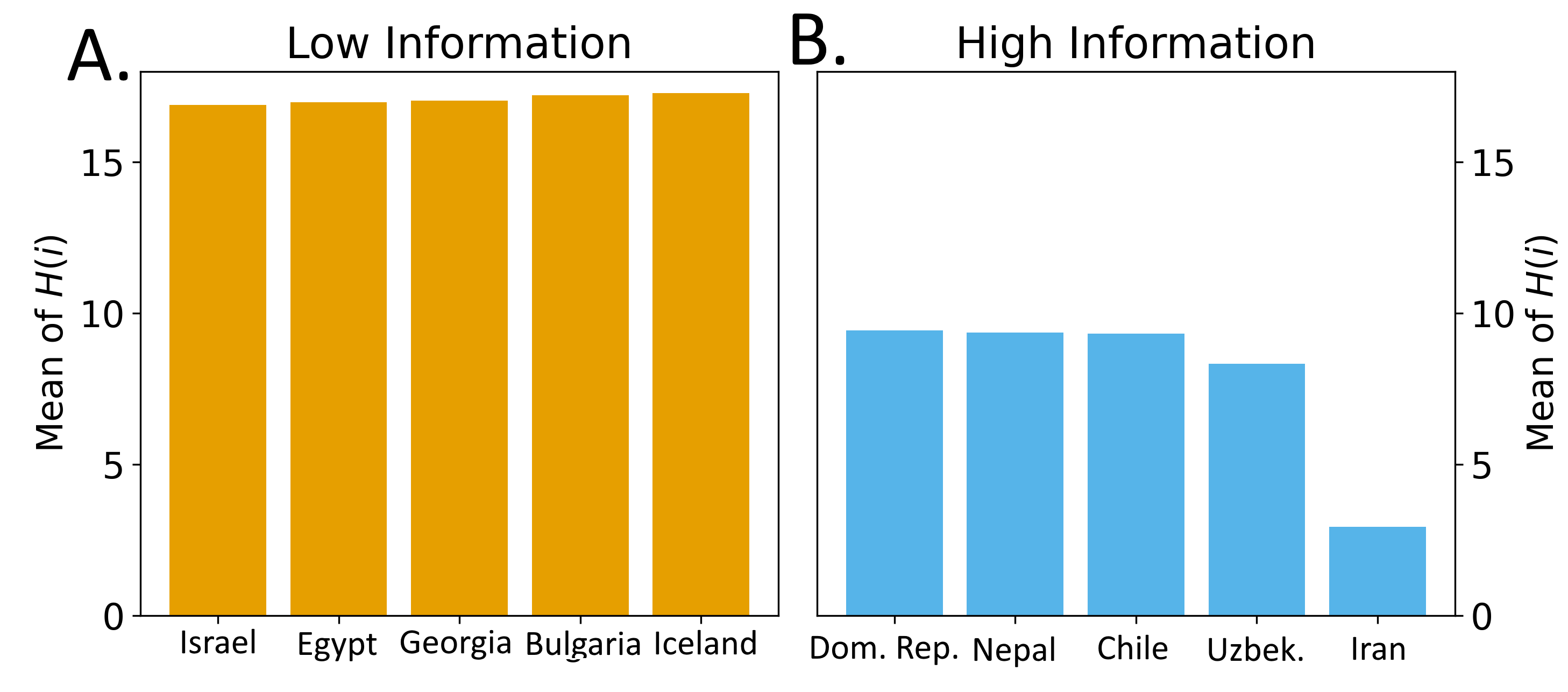}
    \caption{\textbf{AS entropy by nation.}
    The five nations from among all nations with at least 50 registered ASes in which our model has the (\textbf{A}) least certainty and (\textbf{B}) most certainty with respect to the per-nation mean AS entropy.}
    \label{fig:nations}
\end{figure}

Finally, we investigate whether there are geographical regions of the network with low confidence---nations where we would like to see new route collectors and vantage points.
We group ASes according to their registered nations, filter out nations with fewer than 50 ASes, and then rank nations by their mean AS entropy (\figtext~\ref{fig:nations}).
We find that our model is very confident about the structure of the Iranian Internet.
A possible explanation for the model's certainty is that Iran's state-run Internet architecture is accessible through very few gateways, has few edges, and is highly stable~\cite{Leyba2019-bordersgateways}.
The model has the lowest confidence in Israel, Egypt, Georgia, Bulgaria, and Iceland, suggesting that establishing new route collectors or vantage points in these countries would provide valuable measurement data and lead to more reliable reconstructions.

\section{Discussion} \label{sec:discussion}

\subsection{Discussion of Results} \label{subsec:discussresults}

We introduced and validated a Bayesian inference procedure for inferring AS network topology from potentially unreliable route collector data without the need for annotation of ASes, use of heuristics, or other outside information.
Instead, it maximizes the available measurement information to assign probabilities of existence to each edge in the AS topology.

Our comparison of existing network reconstructions highlights the importance of leveraging all available data: the naive reconstructions miss likely edges when excluding IPv6 routes and include unlikely edges by ignoring negative observations.
But publicly available data may not tell the whole story.
The CAIDA AS Relationships reconstructions differ significantly from the others, possibly owing to their inclusion of customer-to-provider peering relationships that do not appear explicitly in public routing tables.
Thus, when using our methods as a tool for evaluating network reconstructions, lower recall (i.e., excluding edges our model believes exist) should cause more concern than lower precision (i.e., including edges our model is skeptical of, possibly owing to a lack of data).

Finally, we showed how our method can be used to identify ASes and national networks for which we lack connectivity information.
The addition of new route collectors increases our model's certainty, though the high normalized entropy indicates our model is realistically cautious about its inference in the face of significant uncertainty in the data---uncertainty that other methods fail to acknowledge.
This corroborates the estimation of Roughan et al.~\cite{Roughan2008-missinglinks} that hundreds more route collectors or vantage points are needed to observe the AS network accurately.
Interestingly, the nations identified by our study to be most in need of more measurement are geographically dispersed.
Many factors may contribute to the measurement data that AS pairs provide in route collector data, such as distance to vantage points and the missing peer problem~\cite{Giotsas2013-missingpeers}.
Identifying the factors that most impact regional information density is an important direction for future work.

\subsection{Optimizations and Implementation Details} \label{subsec:optimize}

The size of the current AS network required several optimizations and implementation decisions related to the general approach described in Section~\ref{sec:methods}.
The May 2021 dataset contains $N \approx \numprint{73000}$ ASes, so the algorithms repeatedly iterated over and stored information about $N(N-1) / 2 \approx \numprint{2.66e9}$ unique AS pairs.
The observation counting algorithm (Algorithm~\ref{alg:obscount}) is easily parallelized to increase runtime efficiency, but requires more effort to store the mapping of AS pairs to their observation vectors tracking the number of time periods each route collector observed each AS pair positively or negatively.
Instead of storing each individual observation vector, we grouped AS pairs into \textit{observation classes}, i.e., sets of AS pairs with the same observation vectors.
This produced a speedup of two orders of magnitude during inference by reducing the set of interest from the \numprint{2.66e9} AS pairs to roughly \numprint{1.5e7} observation classes.

Observation classes enabled additional optimizations during the EM computation to estimate the maximum likelihood parameters $\boldsymbol{\alpha}$, $\boldsymbol{\beta}$, and $\rho$ of our model.
Our parallel EM implementation is optimized both for runtime and memory efficiency, maximizing a value proportional to the log-density of Eq.~\ref{eq:parampost} to improve numerical stability.
Since any edge with the same observation vector contributes the same value to this log-density, we iterate over observation classes instead of all AS pairs, again reducing runtime by two orders of magnitude.
%




\subsection{Limitations} \label{subsec:limitations}

One useful application of reliable AS network reconstructions is in early warning of BGP anomalies, including false route advertisements, prefix hijacking, path poisoning, and BGP black-holing.
There are a variety of existing approaches to this problem. 
Distributed anomaly detection algorithms imbue participating ASes with enhanced security guarantees and increased mitigation time without centralized coordination or fundamental changes to BGP~\cite{Karlin2008-autonomoussecurity}.
Machine learning approaches have been used to identify the signatures of anomalies in BGP time 
series data~\cite{Cheng2018-lstmbgp,McGlynn2019-bgpdeeplearning,Zhang2005-learninganomaly}.
Others search for evidence of misbehaving vantage points using information about the network's structure~\cite{Luckie2014-spuriousroutes}; however, certain substructures appearing in both anomalous and correct data limit the impact of this approach.

In its current form, our Bayesian inference method is not a suitable BGP anomaly detection method despite its ability to infer reliable representations of the AS network topology.
Many BGP anomalies and attacks make use of real edges in which our model tends to have high certainty.
Since edges that are observed positively tend to be observed positively by several route collectors across multiple time periods and rarely have conflicting negative observations, our method struggles to differentiate false announcements from real edges.
Future modifications to the data preprocessing and observation counting methods may enable network inference to serve as an effective anomaly detection scheme.

\section{Related Work} \label{sec:relwork}

\paragraph{BGP Simulation}

Several previous studies have leveraged simulations of BGP routing policies to understand the AS network topology.
Karlin et al.~\cite{Karlin2008-autonomoussecurity} used simulations to validate the security advantages of their ``Pretty Good BGP'' algorithm over the more costly implementation of RPKI.
Gill et al.~\cite{Gill2012-modelingquicksand} performed sensitivity analysis on an ensemble of simulations to infer valley-free routes from established AS business relationships.
\revise{BGP has been simulated to study convergence properties and evaluate the effectiveness of BGP architectures~\cite{Griffin2001-bgpconvergence,Hao2003-bgpsimulation,Dimitropoulos2006-bgpsims}.}
In contrast to simulation, our domain-independent Bayesian inference method directly infers AS-level connectivity from network measurement data, relying on the density of information contained therein.

\paragraph{BGP Topologies}

Previously, researchers have used various inference algorithms to study the structure of the AS topology.
The well-known AS Relationships method~\cite{Luckie2013-asrelationships} \revise{and related methods~\cite{feng2019-unari,Jin2019-problink}} infer business relationships between ASes from public routing data, identifying whether these relationships are peer-to-peer or customer-to-provider.
\revise{Most closely related to our approach, Toposcope~\cite{jin2020-toposcope} employs several methods including Bayesian networks to infer business relationships.
Each of these related works focuses on the business relationship inference problem, which is distinct from the edge existence inference problem. Additionally, these approaches use heuristic filters for selecting paths to use for inference. Prior to this work, the significance of these heuristics has not been explored, but the effect of misbehaving vantage points has been documented~\cite{Luckie2014-spuriousroutes}.}

\paragraph{Network Inference Methods}

Rigorous statistical techniques~\cite{Newman2018-estimatingnetwork,Newman2018-networkstructure,Le2018-estimatingnetwork,Peixoto2018-reconstructingnetworks,Priebe2015-inferenceerrorgraph} have previously been used to determine the structure of empirical networks from noisy data in several application domains including for instance social networks analysis~\cite{Butts2003-networkinference}, the analysis of protein-protein interactions~\cite{Jansen2003-bayesiannetworks}, and ecology~\cite{Young2021-reconstructionplant}.
A closely related line of work focuses on inference with incomplete observations~\cite{Crane2018-statisticalnetwork,Kolaczyk2017-statisticsnetworkanalysis}, investigating what can be learned about a fixed network if its structure is queried one or many paths at a time instead of observed fully.
Crucially, these path-based inference methods assume that whenever an edge is observed in a sample, that edge must exist; i.e., they assume that uncertainty only arises from not seeing the whole network in samples.
In contrast, our approach models measurement as imperfect and thus allows for samples to be affected by error.

\section{Conclusion} \label{sec:conclude}

We introduced a statistical inference method for inferring the AS network topology from noisy, incomplete, path-based observational data.
Our generative model of observational error allows for differing levels of trust across route collectors and was demonstrated to pass established statistical validation tests.
We used \revise{our model} to compare various AS network reconstruction methods and to identify regions of the network where our model had low certainty, suggesting countries that would most benefit from the creation of new route collectors and vantage points.
As global Internet connectivity patterns evolve to reflect ongoing technological, economic, and political changes, we hope that the methods presented here will contribute to more accurate network reconstructions and better-informed interventions.

\revise{The authors have provided public access to their code and data at \url{https://kirtusleyba.github.io/noisynets}.}

\section*{Acknowledgements}

K.G.L.\ and S.F.\ are supported in part by NSF awards CCF-1908233 and IOS-2029696.
J.J.D.\ is supported by the Momental Foundation's Mistletoe Research Fellowship and the ASU Biodesign Institute.
J.G.Y.\ is supported in part by the James S. McDonnell Foundation.
M.E.J.N.\ is supported in part by NSF awards DMS-1710848 and DMS-2005899.

\bibliographystyle{IEEEtran}
\bibliography{ref}

\end{document}